\documentclass[aip,twocolumn,superscriptaddress,reprint]{revtex4-1}

\usepackage{amssymb}
\usepackage{amsmath}
\usepackage{graphicx}
\usepackage{natbib}
\usepackage{epstopdf}

\begin{document}
\textheight = 63\baselineskip

\title{A Reconfigurable Gate Architecture for Si/SiGe Quantum Dots}

\author{D. M. Zajac}
\affiliation{Department of Physics, Princeton University, Princeton, NJ 08544, USA}
\author{T. M. Hazard}
\affiliation{Department of Physics, Princeton University, Princeton, NJ 08544, USA}
\author{X. Mi}
\affiliation{Department of Physics, Princeton University, Princeton, NJ 08544, USA}
\author{K. Wang}
\altaffiliation{Department of Physics, Harvard University, Cambridge, MA 02138, USA}
\affiliation{Department of Physics, Princeton University, Princeton, NJ 08544, USA}
\author{J. R. Petta}
\affiliation{Department of Physics, Princeton University, Princeton, NJ 08544, USA}
\affiliation{Department of Physics, University of California, Santa Barbara, CA 93106, USA}

\date{\today}

\begin{abstract}
We demonstrate a reconfigurable quantum dot gate architecture that incorporates two interchangeable transport channels. One channel is used to form quantum dots and the other is used for charge sensing. The quantum dot transport channel can support either a single or a double quantum dot. We demonstrate few-electron occupation in a single quantum dot and extract charging energies as large as 6.6 meV. Magnetospectroscopy is used to measure valley splittings in the range of 35--70 $\mu$eV. By energizing two additional gates we form a few-electron double quantum dot and demonstrate tunable tunnel coupling at the (1,0) to (0,1) interdot charge transition.
\end{abstract}



\maketitle

Quantum dots have considerable potential for the realization of spin-based quantum devices.\cite{loss1998,ladd2010} Extremely long spin coherence times\cite{muhonen2014,steger2012,saeedi2013} and the ability to utilize existing fabrication processes make silicon an attractive host material for quantum dot qubits.\cite{veldhorst2014,maune2012,dohun2014} Existing depletion mode designs use gate electrode patterns that are much larger than the spatial extent of the resulting electron wavefunctions.\cite{stopa2008} As a result, it is difficult to precisely control the electronic confinement potential. Successful scaling to a larger number of quantum dots will require fine control of the confinement potential on 20 nm length scales. Accumulation mode designs,\cite{angus2007, borselli2014} where electrons are accumulated under small positively biased gates (instead of depleted using large ``stadium" gate designs\cite{petta2005}) allow control of the confinement potential on a much smaller length scale and merit further development.

In this letter we present a reconfigurable accumulation mode device architecture that utilizes three overlapping aluminum gate layers. The device architecture has two parallel (and interchangeable) transport channels. One of the channels is used to create single and double quantum dots, while the other channel is used to define a charge sensor quantum dot.\cite{barthel2010} The natural length scale of this gate architecture is comparable to the resulting dot size, allowing a higher degree of control compared to depletion mode devices.\cite{petta2005} Direct local accumulation also reduces capacitive cross-coupling, simplifying the formation of double quantum dots and tuning of the relevant tunnel rates. The architecture demonstrated here provides a straightforward method for scaling to a larger series array of $N$ quantum dots, with the required number of gate electrodes in each channel growing linearly as 2$N$+1.

\begin{figure}
\includegraphics[width=\columnwidth]{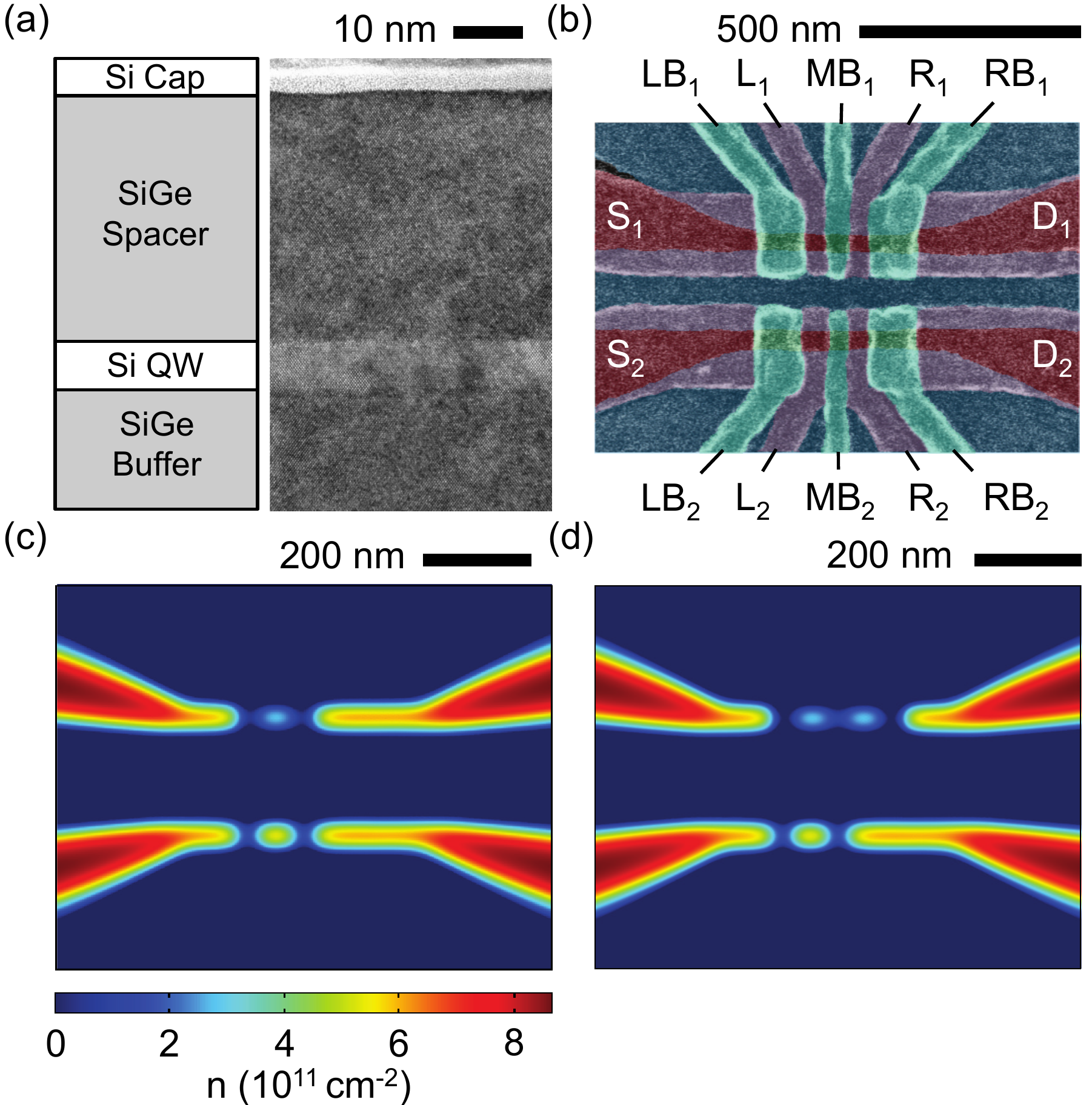}
\caption{\label{Fig1}
(a) Schematic and cross-sectional transmission electron micrograph showing the growth profile of the heterostructure. An 8 nm Si quantum well is grown on top of a 225 nm Si$_{0.7}$Ge$_{0.3}$ layer, followed by a 50 nm Si$_{0.7}$Ge$_{0.3}$ spacer and a 2 nm protective Si cap. (b) False-color scanning electron micrograph of a device identical to the one measured (see text for details). (c) COMSOL simulation of the electron density in the QW with the gate voltages tuned to form a single dot under L$_{1}$ in the upper channel and (d) with a double dot under L$_{1}$ and R$_{1}$  in the upper channel. In (c--d), the lower channel is configured as a charge sensor, with a single dot formed beneath L$_{2}$.}
\end{figure}

The device is fabricated on an undoped Si/SiGe heterostructure with the growth profile shown in Fig.\ \ref{Fig1}(a). A SiGe relaxed buffer substrate is grown on a Si wafer by linearly varying the Ge concentration from 0 to 30\% over 3 $\mu$m. The surface of this virtual substrate is then polished before growing an additional 225 nm thick Si$_{0.7}$Ge$_{0.3}$ layer, followed by an 8 nm Si quantum well (QW), a 50 nm Si$_{0.7}$Ge$_{0.3}$ spacer and a 2 nm protective Si cap. The Si QW is uniaxially strained by the Si/Si$_{0.7}$Ge$_{0.3}$ lattice mismatch, breaking the six-fold valley degeneracy.\cite{ando1982,schaffler1997} The degeneracy of the two lowest lying valleys is further lifted by quantum confinement in the growth direction.\cite{boykin2004} Accumulation mode Hall bar samples fabricated on this wafer yield a two-dimensional electron gas (2DEG) carrier mobility $\mu$ = $1.76 \times 10^5$ cm$^2$/Vs at an electron density $n$ = $2 \times 10^{11}$/cm$^2$ and temperature $T$ = 350 mK. A valley splitting $\Delta_{\rm v}$ = 170 $\mu$eV is measured at a magnetic field $B$ = 1.75 T. The 2DEG also undergoes a metal-to-insulator transition (MIT) at a critical density of $n_{\rm c}=0.46 \times 10^{11}$/cm$^2$. Both the high electron mobility and low critical density of the MIT are indicative of low disorder in the 2DEG.\cite{xiao2015}

Electronic confinement in the plane of the QW is achieved using three overlapping layers of Al gate electrodes, as shown in Fig.\ \ref{Fig1}(b). Overlapping gates allow full control of the confinement potential since no region of the Si surface is left exposed. In addition, the potential is tuned on a 20 nm length scale, limited by the resolution of the electron beam lithography tool. The first layer of aluminum, shown in light blue, serves as a screening layer, and its purpose is to selectively screen out the electric fields formed by the accumulation gates in layers 2 and 3. The result is two parallel transport channels that cross the device from left to right. The second Al layer, shown in red, consists of two plunger gate electrodes in the upper channel (labelled L$_{1}$ and R$_{1}$) and lower channel (labelled L$_{2}$ and R$_{2}$). Source and drain accumulation gates (S$_{1,2}$ and D$_{1,2}$, respectively) are also defined in layer 2 and are typically biased at 700 mV to create a Fermi sea of electrons that extends out to $n^{++}$ implanted ohmic contacts. Layer 3, shown in green, consists of three gate electrodes per channel, labelled LB$_{1,2}$, MB$_{1,2}$, and RB$_{1,2}$, which are designed to control tunnel barrier heights. A controlled oxidation process is performed after the metallization of each layer to create a thin insulating oxide around the gates, which electrically isolates them from subsequent gate layers.\cite{lim2009,lim0929}

Given the symmetry of the design, the upper and lower transport channels are interchangeable. In what follows, we utilize the lower channel to form a charge sensor quantum dot that is sensitive to the electron occupancy of the quantum dots formed in the upper channel. To form the sensor quantum dot we first raise the voltages on all gates in the lower channel until conductance is measured through the channel. The gate voltages $V_{\text{LB}_2}$ and $V_{\text{MB}_2}$ are then lowered to form tunnel barriers, resulting in a single quantum dot under L$_{2}$. A single quantum dot is formed in the upper channel by performing the same procedure, resulting in a single quantum dot under L$_{1}$. $V_{\text{L}_1}$ is then decreased until we achieve few-electron occupancy. The carrier density for this gate voltage configuration is simulated using COMSOL Multiphysics software,\cite{schmidt2014} as shown in Fig.\ \ref{Fig1}(c). In principle, an additional quantum dot can be formed under R$_{1}$ by lowering $V_{\text{RB}_1}$. Again $V_{\text{R}_1}$ is decreased to reduce the electron occupation in the right dot resulting in the simulated carrier density shown in Fig.\ \ref{Fig1}(d).

\begin{figure}
\includegraphics[width=\columnwidth]{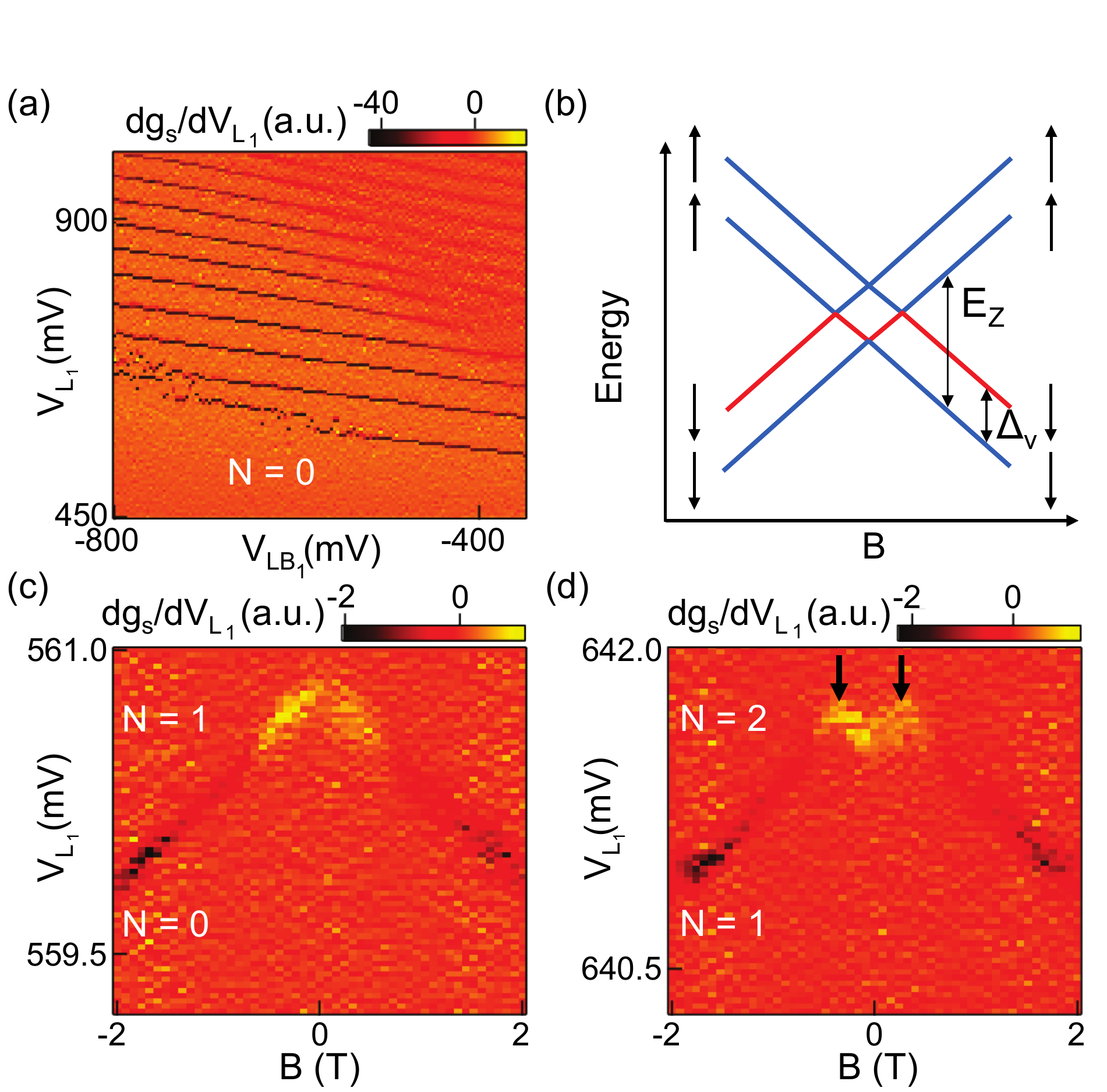}
\caption{\label{Fig2}
(a) The derivative of the charge sensor conductance $dg_\text{s}/dV_{\text{L}_1}$ plotted as a function of $V_{\text{L}_1}$ and $V_{\text{LB}_1}$ reveals the charge stability diagram of the single quantum dot formed under L$_{1}$. (b) Energy level diagram of a quantum dot with a low-lying valley state plotted as a function of $B$, where $E_\text{z}$ is the Zeeman energy and $\Delta_\text{v}$ is the valley splitting. The red line shows the expected spin filling of the $N$ = 1 to 2 charge transition as a function of $B$. (c) Magnetospectroscopy at the $N$ = 0 to 1 charge transition is consistent with spin-down filling. (d) Magnetospectroscopy at the $N$ = 1 to 2 charge transition indicates a crossover from spin-up filling to spin-down filling when $E_\text{z}$ = $\Delta_\text{v}$.}
\end{figure}

We now show that this gate architecture is capable of reaching the few electron regime in both single and double dot modes of operation. The device is characterized in a dilution refrigerator with an electron temperature $T_{\rm e}$ = 40 mK. We first form a single lead quantum dot in the upper channel using gate L$_1$ as a plunger and gate LB$_1$ to control the tunnel barrier height. Charge sensing is achieved by forming a single dot in the lower channel under L$_2$ using LB$_2$ and MB$_2$ as barriers. The charge sensor quantum dot is biased on the edge of a Coulomb blockade peak, where the dot conductance is highly sensitive to changes in the local electrostatic potential.\cite{elzerman2003} Figure \ref{Fig2}(a) shows a charge stability diagram obtained by plotting the derivative of the charge sensor conductance d$g_{\rm s}/$d$V_\text{L$_1$}$ as a function of the voltages $V_{\text{L}_1}$ and $V_{\text{LB}_1}$. $V_{\text{L}_2}$ is linearly compensated as $V_{\text{L}_1}$ and $V_{\text{LB}_1}$ are swept to keep the dot biased on the edge of the Coulomb blockade peak. The absence of charge transitions in the lower left corner of the plot indicates that we have reached the $N$ = 0 charge state.

Tight electronic confinement is important in Si quantum devices since the effective mass is approximately three times larger than in GaAs. We can estimate the size of the accumulation mode quantum dots from measurements of the charging energy. The spacing between the first two charge transitions in Fig.\ 2(a) is $\Delta V_{\text{L}_1}$ = 60 mV. Taking into account the lever-arm conversion between gate voltage and energy $\alpha=0.11$ eV/V$_{\text{L}_1}$, this yields a charging energy of 6.6 meV. We estimate the dot size from the capacitance using a disk capacitor model. From the charging energy we extract a capacitance of $C=e^2/E_{\rm c}= 24.3$ aF where $E_{\rm c}$ is the charging energy,\cite{kouwenhoven1997} giving a dot radius of $r=C/8\epsilon_{\rm r}\epsilon_0= 29$ nm where $\epsilon_{\rm r}=11.7$ and $\epsilon_0=8.85 \times 10^{-12}$ F/m. This is consistent with the dot radius of $\sim$ 30 nm that is predicted using COMSOL. Pulsed-gate spectroscopy data\cite{elzerman2004} (not shown) yields an orbital energy $E_{\rm orb}$ = 2.5 meV for the $N$ = 1 charge state, which is also consistent with a dot size of $r$ = 29 nm for a particle in a 2D box. By setting the area of the disk equal to the area of a 2D square box with side $L$ ($\pi r^2$ = $L^2$), we expect an orbital energy of $E_{\rm orb}=3\hbar^2\pi^2/2m^{*}L^2=2.2$ meV where $\hbar$ is the reduced Planck's constant, $m^{*}$ = 0.19 $m_{\rm e}$, and $m_{\rm e}$ is the free electron mass.\cite{schaffler1997} Lastly, Fig.\ 2(a) indicates that $V_{\text{LB}_1}$ is very effective at tuning the tunnel rate to the source lead of the dot. For example, the $N$ = 0 to 1 and $N$ = 1 to 2 charge transitions show ``latching" in the lower left region of the plot when the barrier tunneling rate $\Gamma$ becomes comparable to the rate at which the gate voltage is rastered. Transitions in the upper right region of the charge stability diagram become lifetime broadened when $\Gamma$ $\sim$ $k_\text{B}T/h$, here $k_\text{B}$ is Boltzmann's constant and $h$ is Planck's constant.\cite{de2001}

Unlifted valley degeneracy will introduce orbital decoherence and weaken Pauli blockade, which is used for spin readout in spin-to-charge conversion.\cite{elzerman2004} It is therefore important to understand the level structure of the quantum dots. Figure \ref{Fig2}(b) schematically shows the energy level diagram of a single quantum dot as a function of $B$ for a relatively small $\Delta_{\text{v}}$.\cite{lim2011} At $B$ = 0 there are two spin degenerate valleys separated by $\Delta_{\text{v}}$. Application of a magnetic field results in Zeeman splitting $E_\text{z}$=$g\mu_\text{B} B$, lifting the spin degeneracy. Here $g$ is the g-factor of electrons in silicon and $\mu_\text{B}$ is the Bohr magneton. Electrons will fill the lowest energy states as they are added to the dot. The red line shows the expected filling of the second electron as a function of $B$. The kink represents the point at which $E_{\text{z}}$ = $\Delta_{\text{v}}$, where the spin state of the loaded electron is expected to change from spin-up to spin-down. Figures \ref{Fig2}(c) and (d) show magnetospectroscopy\cite{borselli2011, yang2012} data taken at the $N$ = 0 to 1 and $N$ = 1 to 2 transitions in the dot formed under L$_1$. The derivative of the charge sensor conductance d$g_{\rm s}/$d$V_{\text{L}_1}$ is plotted as a function of $V_{\text{L}_1}$ and $B$. The presence of a kink in Fig.\ \ref{Fig2}(d) is consistent with the red line shown in Fig.\ \ref{Fig2}(b) indicating a change from spin-up filling to spin-down filling at $B=0.35$ T. From the lever arm $\alpha=0.11$ eV/V$_{\text{L}_1}$ and the slope of the charge transition in Fig.\ \ref{Fig2}(c) we measure a g-factor of $g$ =1.9 $\pm$ 0.1 and extract a valley splitting of $\Delta_{\text{v}}$ = 35 $\mu$eV for this dot. Single lead quantum dots were also formed under gates R$_1$, L$_2$, and R$_2$ by reconfiguring the device (data not shown here) yielding valley splittings of 35 $\mu$eV, 60 $\mu$eV, and 70 $\mu$eV.

\begin{figure}
\includegraphics[width=\columnwidth]{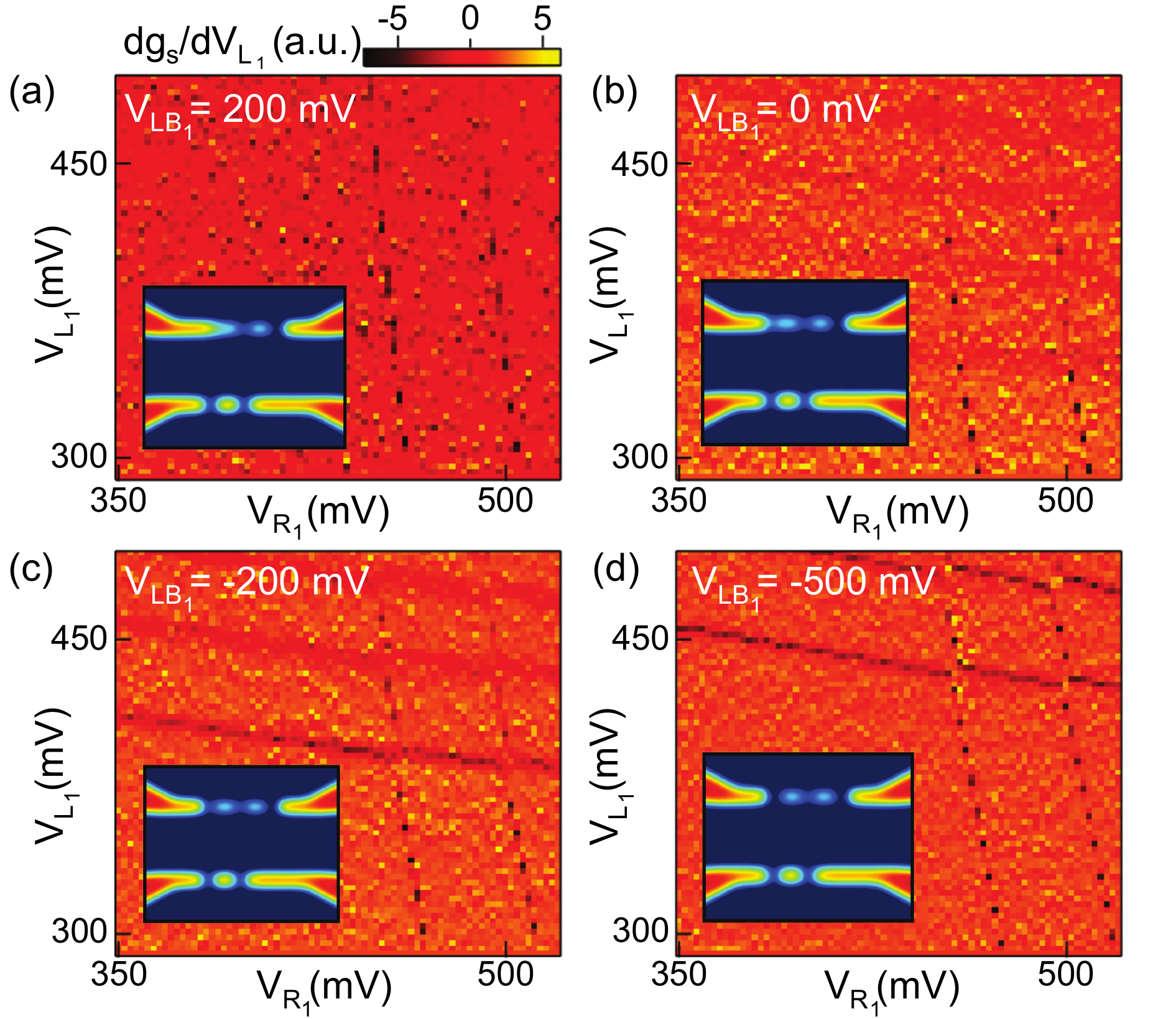}
\caption{\label{Fig3}
Evolution of the charge stability diagram as a function of $V_{\text{LB}_1}$. Charge stability diagrams obtained by sweeping $V_{\text{L}_1}$ and $V_{\text{R}_1}$ are shown for: (a) $V_{\text{LB}_1}$ = 200 mV, (b) $V_{\text{LB}_1}$ = 0 mV, (c) $V_{\text{LB}_1}$ = -200 mV, and (d) $V_{\text{LB}_1}$ = -500 mV. As $V_{\text{LB}_1}$ decreases a new dot is formed under L$_1$, converting the single quantum dot into a double quantum dot. Insets show COMSOL simulations of the electron density in the QW.
}
\end{figure}

Device tuning requires control of a multi-dimensional parameter space.\cite{granger2010} In depletion mode devices there is significant cross-capacitance between gates\cite{hanson2007} which complicates this task.  The cross capacitance of adjacent gates can be extracted from the slope of the charge transitions in single dot charge stability diagrams. For example, the data of Fig.\ \ref{Fig2}(a) yield a capacitance ratio of 19\%, in reasonable agreement with COMSOL predictions (12\% cross capacitance for these gates). The small cross-coupling allows a single dot in the upper channel to be easily converted into a double quantum dot by changing a single gate voltage.

Figure \ref{Fig3} (a--d) show charge stability diagrams that are obtained by sweeping $V_{\text{L}_1}$ and $V_{\text{R}_1}$. In Fig.\ \ref{Fig3}(a) $V_{\text{LB}_1}$ = 200 mV and a single dot is formed under the right plunger gate, R$_1$. A tunnel barrier is then formed under LB$_1$ as its voltage is made more negative, converting the single quantum dot into a double quantum dot. At $V_{\text{LB}_1}$ = 0 mV a dot begins to form under L$_1$ as indicated by the appearance of lifetime broadened charge transitions, see Fig.\ \ref{Fig3}(b). Further reducing $V_{\text{LB}_1}$ strengthens Coulomb blockade, resulting in sharper left dot charge transitions. A clean double dot stability diagram is obtained in Fig.\ \ref{Fig3}(d) for $V_{\text{LB}_1}$ = -500 mV. Insets in each panel show the calculated electron densities for different values of $V_{\text{LB}_1}$ and serve to illustrate the evolution of the 2DEG at different stages of tuning.

\begin{figure}[t]
\includegraphics[width=\columnwidth]{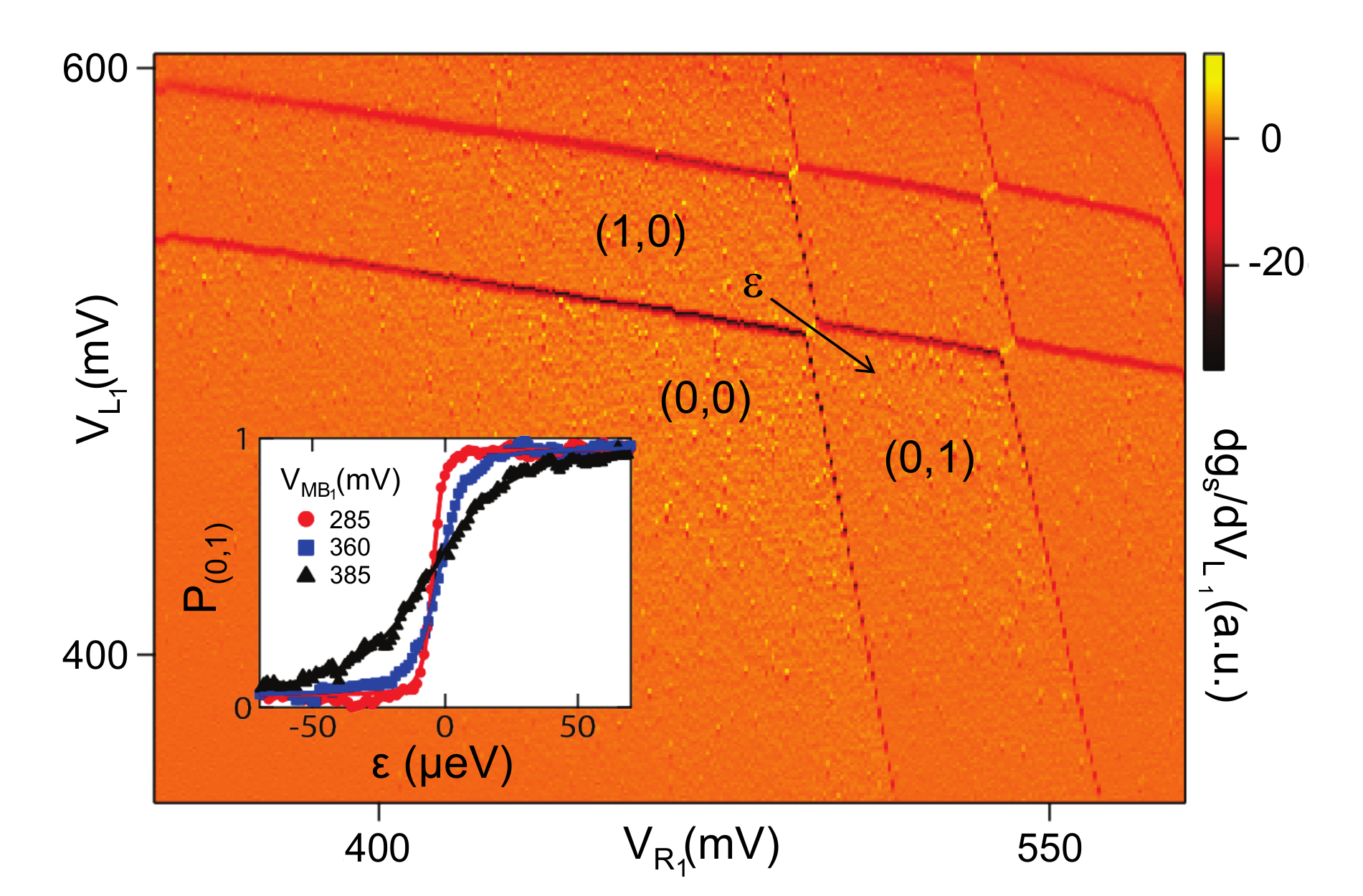}
\caption{\label{Fig4}
Double quantum dot charge stability diagram showing few-electron occupation. Inset: The probability of being in the (0,1) charge configuration $P_{\text{(0,1)}}$ is plotted as a function of detuning $\epsilon$ at the (1,0)-(0,1) charge transition for different values of $V_{\text{MB}_1}$ showing tunable interdot tunnel coupling.
}
\end{figure}

Figure \ref{Fig4} shows a high resolution double dot charge stability diagram. Here $V_{\text{L}_2}$ is linearly compensated to stay on the edge of a Coulomb blockade peak while $V_{\text{L}_1}$ and $V_{\text{R}_1}$ are swept. The absence of charge transitions in the lower left corner of the plot indicates that we have reached the (0,0) charge configuration, where ($N_L$,$N_R$) denotes the occupancy of the left and right dots. The left and right dot charge transitions show very little curvature, indicating that the confinement potential is well defined by our gate pattern and not significantly affected by the capacitive coupling of the two dots. From the slope of a left dot charge transition we extract a right plunger to left plunger capacitance ratio of 24\%, which is considerably smaller than the 55\% coupling measured in a dual-gate architecture.\cite{wang2013}


The ability to tune the tunnel coupling between two dots is an important requirement for spin-based quantum dot qubits.\cite{divincenzo2000} First, it demonstrates control of the confinement potential on short length scales. Second, the interdot tunnel coupling sets the double dot exchange energy $J=4t_{\rm c}^2/E_{\rm c}$\cite{loss1998}. The Hamiltonian at the (1,0)-(0,1) charge transition is given by $H=(\epsilon /2)\sigma_{\rm z} + t_{\rm c}\sigma_{\rm x}$ where $\epsilon$ is the detuning, $t_{\rm c}$ is the interdot tunnel coupling and $\sigma_{\rm i}$ are the Pauli matrices.\cite{kouwenhoven2002} We demonstrate tunable interdot tunnel coupling in the inset of Fig.\ \ref{Fig4}, which shows the probability of being in the (0,1) charge state $P_\text{(0,1)}$ as a function of $\epsilon$ for three different values of $V_{\text{MB}_1}$. The detuning axis is shown in the main panel of Fig.\ \ref{Fig4}. These data are fit to the expression
\begin{equation}
P_{(0,1)}=\frac{1}{2}\left[ 1+\frac{\epsilon}{\Omega} \text{tanh}\left(\frac{\Omega}{2k_\text{B}T_{\rm e}}\right)\right],
\end{equation}
where $T_{\rm e}\sim$ 40 mK is the electron temperature and $\Omega=\sqrt{\epsilon^2+4t_{\rm c}^2}$ is the energy difference of the hybridized charge states.\cite{dicarlo2004, petta2004, simmons2009} For $V_{\text{MB}_1}$ = 285 mV, $2t_{\rm c}<k_{\text{B}}T_{\rm e}$ and the transition is thermally broadened. For $V_{\text{MB}_1}$ = 360 and 385 mV values of $t_{\rm c}$= 5 $\mu$eV and 15 $\mu$eV are extracted by fitting the data to Eq.\ (1). For higher values of $V_{\text{MB}_1}$ we observe tunnel couplings exceeding 100 $\mu$eV.

In conclusion we have demonstrated a reconfigurable device architecture for Si/SiGe that allows the formation of single and double quantum dots. In single dot mode, we extract charging energies as large as 6.6 meV, orbital energies of 2.5 meV, and valley splittings in the range of 35 -- 70 $\mu$eV. With the same device, we have formed a few electron double quantum dot, demonstrating tunable interdot tunnel coupling. The overlapping gate architecture provides a path forward for scaling to larger series arrays of quantum dots.

\begin{acknowledgements}
Research was sponsored
by the United States Department of Defense with
partial support from the NSF (DMR-1409556 and DMR-1420541). The views and conclusions
contained in this letter are those of the authors
and should not be interpreted as representing the official
policies, either expressly or implied, of the United States
Department of Defense or the U.S. Government.
\end{acknowledgements}

\end{document}